\documentclass[a4paper,20pt]{article}
    \usepackage[T1]{fontenc}
    \usepackage{graphicx}
    \usepackage{epsfig}
    \usepackage{amsmath}
    \usepackage{rotating}
    \usepackage{amsfonts}
    \renewcommand{\abstract}{}
\newcommand{\solar}{$_{_{\bigodot}}$}

\begin{document}
\makeatletter
\renewcommand{\@oddhead}{\textit{YSC'14 Proceedings of Contributed Papers} \hfil \textit{A. Al-Sawad}}
\renewcommand{\@evenfoot}{\hfil \thepage \hfil}
\renewcommand{\@oddfoot}{\hfil \thepage \hfil}
\fontsize{11}{11} \selectfont

\title{Multi Eruption Solar Energetic Particle Events Observed with SOHO/ERNE}
\author{\textsl{Amjad Al-Sawad$^{1}$}}
\date{}
\maketitle
\begin{center} {\small $^{1}$Department of Physics and the V\"{a}is\"{a}l\"{a} Institute for
Space Physics and Astronomy,\\ Turku University, FIN 20014 Turku,
Finland\\ amjal@srl.utu.fi}
\end{center}

\begin{abstract}
A combination of many Solar energetic particle (SEP) events, each
one of which is associated with a single eruption, can create one
complex intensity-time profile, that will result in masking the
observation of the first injected particles detected near Earth for
each participated eruption. We defined such SEP events as Multi
Eruption Solar Energetic Particle (MESEP) events. We have
investigated the intensity-time profile of 333 solar energetic
particle events during the operation time of SOHO mission and
studied the associative solar eruptions (CMEs and solar flare) from
the starting time of each event till the end. We found that most of
the events have multi eruption phenomena which might or might not
affect the intensity-time profile. We found that it is possible to
know the real effect of some of the eruptions during the whole
duration of the event, even if their effect as masked by the first
eruption, by studying the widest possible energy range, the $^4He/P$
ratio and the anisotropy.
\end{abstract}

\section*{Introduction}
\indent \indent Since the 1970's Solar energetic particle (SEP
hereafter) events have been well known to be connected with two
major coronal phenomena: solar flares and Coronal Mass Injections
(CMEs hereafter) (e.g., \cite{can86, kah77, she75, she83a}), and
with interplanetary shocks \cite{she83b,she85}. The observation of
the first non-scattered relativistic particles \cite{tor99} is one
tool to establish the connection between sudden rise in the
intensity-time profile for many particle species which occur after
flares or CMEs. Note that these measurements are background
dependent and unless the background of the intensity-time profile is
clear enough and not masked by previous events, the exact
association of the SEP events with the eruption cannot be accurate.
In this manner it is not possible to know whether the CMEs or solar
flare which occurred after the first eruption have
injected/accelerated any SEPs since intensity-time profiles are
masking the possible effect of the SEPs related to such eruptions.
Furthermore, even in some single eruptive SEP events we can find
more than one acceleration phase (e.g., \cite{Al06, bom06}). Cane et
al. \cite{can02} found correlation between long-low frequency type
III radio bursts and SEP events. During one event we might observe
many such radio emissions and hence multi-eruption sources cannot be
ruled out.

However, many earlier studies indicate the general features of CMEs
and solar flares which are associated with SEP events whether they
are impulsive or gradual (e.g., \cite{cli96}). It is well known that
the CME speed and the SEP intensity are well correlated
\cite{kah01}. The same study concludes that no fast CMEs with widths
less than $60^\circ$ are associated with SEP events. And nearly all
fast halo CMEs are associated with SEP events. On the other hand,
the indication of the solar flare X-ray features related to SEP
events has been established earlier (e.g, \cite{can86}). In general
we know now that CMEs with speed $>$500 km/s and angular width
$>50^{\circ}$ can be possibly related to injection/acceleration of
SEP and long duration C, M, X class solar flare or impulsive M and X
class can also be related to injection/acceleration of SEP.

A gradual SEP event might not be due to one single eruption, CME
or solar flare. The continuity of high intensities due to coronal
or interplanetary shock waves might come from several eruptions,
which might end in showing one single prolonged intensity-time
profile. In such case a continuing intensity-time profile can be
due to Multi Eruption SEP (MESEP) event. Whether those eruptions
are effective or not this issue needs further examination. Note
that the starting of an event might be due to a CME associated
with a solar flare. Due to complexity of distinguishing between
CMEs and flares in particle acceleration is a non-solved problem,
since both eruptions occur in general in same active region and
same time, we rather account such kind of events as single
eruptive.

In several studies (e.g., \cite{can03}) the ratio of different
compositions has revealed the differences in seed population. Cliver
\cite{cli96} and Reames \cite{rea99} have used $^4He/P$ ratio to
identify different classes of SEP events. Thus the changing in the
$^4He/P$ ratio during the specific period might be related to
different tubes of SEPs accelerated by different eruptions. On the
other hand, studying the history of particle transport might reveal
the different sources for SEP acceleration. Moreover, Cliver et al.
\cite{cli04} indicate that changes in SEP time profile is due to
more than one SEP-effective shock or other accelerator at work.

In this study we examine (SEP) events which have an intensity-time
profile masking the effect of later eruptions in the event. The
examination of possibly widest range of energy channels with the
investigation of $^4He/P$ ratio and anisotropy might help to
determine whether those eruptions are effective or not.

\section*{Data analysis}
\indent \indent We have used intensity-time profile for proton and
helium particles provided by the two ERNE detectors, High Energy
Detector (HED) and Low Energy Detector (LED) (Torsti et al.
\cite{tor97}). We examine the intensity-time profile through two
ways.\\
I) The linear profile provided by
${http://www.decent.fi/soho/index.php}$.\\ II) The logarithmic
profile provided by ${http://www.srl.utu.fi/erne\_data/}$. \\ In the
linear fit the changes during high-intensity periods are more
visible than in logarithmic view and thus it might be easier to see
if the changes were due to local effect or they have a velocity
dispersion and belong to an event on the Sun. Logarithmic scale is
used by most researchers and gives the general view of the event.

We took the widest possible energy range starting from 1 MeV up to
116 MeV for each event to see the differences in the profile in each
channel and whether it is the same for all events or we have
different individual cases or groups. We considered an event to be a
MESEP event if the logarithmic proton intensity-time profile does
not reach the background level during a certain period in many
energy levels (some might do) and we observed many eruptions on the
Sun during that period. We divided the MESEP events according to the
energy levels to three types 1) low energy events (weak events)
which produce particle of $<$10 Mev and with maximum intensity of
not less than 10$^{-2}$/(cm$^2$~sr~s~MeV). 2) Mid energy MESEP
events, which produce particles of $<$30 MeV. 3) High energy MESEP
events, which produce particles of $>$30 MeV, and to four types
according to the intensity-time profile in the low and high
energies. Type 1) where the low-energy channels show peaks and the
peaks fade down or disappear in the higher energy channels leading
to less peaks or a single peak. Type 2) where both low and high
energy channels show the peaks. Type 3) where the high energy
channels show peaks that low energy channels does not show. Type 4)
where both low and high energy channels does not show any obvious
peaks (see Fig.1).

We used the SOHO/LASCO catalog at
${http://cdaw.gsfc.nasa.gov/CME\_list/UNIVERSAL/}$ to determine the
number of CMEs related to those events, taking into consideration
CMEs with $>$500~km/s and GOES data search engine for solar flare,
taking in consideration M, X and long duration C class flares. We
have selected some events and look for: (i) Any changes in any
energy channel close in time to the eruption, especially if the
changing create peaks that shows a rise intensity closer to the
background like in the beginning of the event. (ii) The measurements
for the $^4He/P$ ratio and comparison of it to the time location of
the eruption. The boundary of ratio changing before and after the
time of the associated eruption should not be less than factor of
two. (iii) Possible anisotropy measurement and compare the changing
to the associated eruption. If any of the above measurements fit
with any eruption we can assume that this eruption has a SEP
enhancement during the event. We also look at the stream
interactione region events list by Jian et al. \cite{jia06a} and the
interplanetary coronal mass ejection list by Jian et al.
\cite{jia06b} to find possible association with these two types of
events. In the future we intend to verify this method and apply it
to all events.

\section*{Results}
\indent \indent We went through the SEP events' intensity-time
profiles from May 1996 till March 2007. A set of 333 SEP events
observed by ERNE was uploaded from the web list
${http://www.decent.fi/soho/index.php}$. We searched in each event
for associated CMEs and solar flares during the time when the
intensities are prolonged and make one continued profile in many
energy channels. Events typically start with a sudden rise in the
intensity over the cosmic ray background due to an eruption launched
on the Sun and the majority of the SEPs start to be injected in the
corona at heights below $\sim$2 R\solar \cite{koc02}. The bulk of
ions of any energy is accelerated by the bow shock of the CME while
it travels through the high corona ($>$5 R\solar \cite{kah94}).
Thus, the first associated eruption can be easily detected with
observation of the first non-scattered energetic particles
independent of acceleration mechanism/release height.

In many events when the acceleration is prolonged for many days,
especially during the maximum activity in the Sun, we observe
frequent eruptions each day. We might not be able to see the first
sudden rise in the intensity nor even the decay. Among the 333
events 268 were observed with many eruptions that associated but not
necessary participated in one intensity-time profile. We excluded
events that are either clearly due to only one solar eruption (in
addition to those with no solar signatures).

Among 65 non-MESEP events about 47 (72\%) were low energy events.
Among 268 MESEP events only 30 (11\%) were low energy events,
leaving a conclusion that the MESEP events are mostly powerful long
duration events. Thus we put the MESEP events according to the
maximum accelerated energy particles into three types: High, Mid and
Low. The major portion ($\sim$70\%) of the MESEP events  were High
energy events - 189 from 268 with energies $>$30MeV. The Mid energy
events 20-30 MeV were 49 from 268 events ($\sim$19\%) and the Low
energy events $<$20 MeV were only 30 events ($\sim$11\%).

The intensity-time profiles are not the same in all the cases of the
MESEP events. It is possible to notice a few types of common
features among them. We have chosen four types (see Fig.1) according
to the following common features in the intensity-time profile for
the 268 MESEP. Type 1, where we normally observe in the low energy
channels many peaks which might or might not be related to events on
the Sun or in the interplanetary medium, but in the high energy
channels show only a few single peaks or even one peak. This means
that some eruptions which are contributing to the events produce
less higher energetic particles than those which maintain the
production or the events have prolonged intensity profile in the low
energy, which is typical. This type was the most common among the
MESEP intensity profile, $\sim$40\% (109 from 268). Type 2, where we
observe in both high energy channels and low energy channels almost
the same changing in the intensity profile. This type contains
$\sim$21\% (56 from 268) of the MESEP events. Type 3, where we
observe many peaks in the high energy channels but not in the low
ones. This type is less common among the MESEP events and contain
only $\sim$16\% (42 from 268). Type 4, where we can not see any
obvious peaks in both high and low energy channels. This type
contain $\sim$23\% (61 from 268) of the MESEP events.

In the first three types it is possible to figure out some of the
hidden eruptions under the prolonged intensity profile if we take
all the possible observed energy channels, since some channels,
either low ones or high or both, have some obvious peaks that might
relate to those hidden eruptions. In those cases we can assume that
the eruptions are effective and have accelerated SEPs (see Fig.1).
But also in certain periods during the events we might observe many
eruptions that do not have any related features in the
intensity-time profile. Those eruptions and the eruptions associated
with type 4, where no peaks are observed, are the most obscure cases
in the SEP acceleration. If we can not see related accelerated
particles through the intensity-time profile then how can we know
whether this eruption has injected or accelerated any SEPs?

We have chosen arbitrary events where we have an analysis of
anisotropy data. An anisotropic event was registered on the 11th of
November starting at 9:45 with the Energetic and Relativistic Nuclei
and Electron (ERNE) instrument on the Solar and Heliospheric
Observatory (SOHO) (see Fig.2). The high energy detector (HED) of
the ERNE instrument is pointing to the direction of
$\theta$=0$^\circ$, $\phi$=315$^\circ$ GSE, and its wide viewcone
(120$^\circ$x120$^\circ$) is divided into 241 directional bins, from
which proton and helium fluxes can be measured (see also Torsti et
al. \cite{tor04}). The upper left panel in Fig.2 shows the
directional distribution at 9:51-9:59 UT for protons from energy
range 16.9-22.4 MeV. The upper right panel shows the pitch angle
distribution the directional bins form. SOHO does not have a
magnetometer on board, but the supposed magnetic field direction was
found by fitting a polynomial function to the pitch angle
distribution (see also \cite{tor06}). The direction of the found
symmetry axis is pointed with a white cross and the deflection
angles of 5$^\circ$, 30$^\circ$, 60$^\circ$ and 90$^\circ$ are
marked with circles. The anisotropic flow is very narrow and
anisotropy merges without significant change in the total mean flux.
The horizontal panel shows an arbitrary statistical anisotropy index
rising to its peak at 10 UT and after a more isotropic period it
rises again. The whole event lasts for less than four hours.

The anisotropic flow of solar energetic protons of 16.9-22.4 MeV at
its peak at 9:51-9:59 UT was measured by ERNE/HED instrument. Time
integration of 8 minutes was used. Upper left is the instrument's
viewcone with coordinates in GSE. The direction of the Sun is
indicated with a small picture of a Sun left from the center of the
viewcone. The full circle area with coordinate lines is the
hemisphere which ERNE is pointing at, and the semi-rectangular
borders indicate the borders of the viewcone. The same 241 measuring
points seen on the left form the pitch angle distribution in the
right panel. The direction that has been used to derive the pitch
angles is indicated in the upper right corner of the panel. The
horizontal panel is the total mean anisotropy index in which values
over 10 indicate significant deviation from isotropic flow.

On the other hand, the measurements of the $^{4}$He/p ratio shows a
clear dip at the same time of the anisotropy changes (see Fig.2).
The $^{4}$He/p can strengthen the assumption for a source of
eruption from the Sun. In some cases anisotropy measurements by
itself might indicate an influence of entering a different magnetic
tube due to the same eruption that fills many magnetics tubes. On
the other hand, the $^{4}$He/p ratio means that we have a source of
different seed population and there are different accelerated
particles at that time. This might lead us to conclude that the
eruption which is associated with the changing in the anisotropy
measurements and $^{4}$He/p has fill a new magnetic tube with
different seed population particles than the previous one. The
eruption in this case was the Lasco CME which has been seen by Lasco
C2 at first time on height 3.51 R\solar with linear speed of 639
km/s and angular width of 34$^\circ$ which is not wide enough as it
is expected for a CME to accelerate SEPs. But the location of the
CME on the solar disc at the southwest in central position angle of
259$^\circ$, can make this CME a good candidate. Still this case can
not be assumed as an effective CME unless we take further careful
analysis. Jian et al. \cite{jia06b} have used the data from Wind and
ACE. His list shows that the space craft has entered an
interplanetary coronal mass ejection at 4:13 UT and ended at 21:00.
There might be a possible existence of different tube which carries
different seed population of energetic particles. In this case the
change in the anisotropy is spatial and due to local effect, but the
type of changing in both $^{4}$He/p ratio and anisotropy suggests
temporal effect from a source on the Sun. We expect to obtain this
method for further investigation for the rest of the hidden
eruptions of the MESEP events in the future.

\section*{Conclusions}
\indent \indent A combination of many SEP events, each one awhich is
ssociated with a single eruption, can create one complex
intensity-time profile that will result in masking the observation
of the first injected particles detected near the Earth for each
participated eruption. We defined such SEP events as Multi Eruption
Solar Energetic Particle (MESEP) events. We found among the 333 SEP
events that we observed between May 1996- March 2007 the following:
\begin{enumerate} \item [1-] About $\sim$80\% of the SEP events were MESEP events.
\item [2-] The MESEP events' intensity-time profiles are observed
with energetic particles from 1 up to over 100 MeV and contain
three types of energy levels: High energy events with energies of
$>$30MeV and $\sim$70\% from total MESEP events; Mid energy events
with energies of $>$20MeV-$<$30MeV and $\sim$19\%; and Low energy
events with energies of $<$20MeV and $\sim$11\%, which means that
the MESEP events are mostly long duration high energy events.
\item [3-] The intensity-time profile for the MESEP are of four
types. Type 1, with peaks in low energy channels disappearing or
getting less in high energy channels and this type contains
$\sim$40\% of the MESEP events. Type 2, with peaks in both high
and low energy channels and contains $\sim$21\%. Type 3, with
peaks in high energy channels more than in low energy channels and
contains $\sim$16\%. Type 4, with no peaks in both energy channels
and contains $\sim$23\%. \item [4-] The observation for the
possible particle acceleration or injection of each participant
eruption are possibly accomplished if we use: i) High range of
intensity-time energy channels which might result in finding some
associated features in some energy channels that might related to
certain eruptions. This is possible in the first three types. ii)
For the rest of the eruptions in type 4 and eruptions in the other
types which are not associated with peaks in any energy channels
we are not able to assume whether they participate in the
acceleration or injection of energetic particles during the period
of the event unless we analyzed the anisotropy and the $^4He/P$
ratio and find obvious association with the eruption. Hidden
eruptions during the MESEP events need further investigation to
provide wider and clearer view on the solar energetic particles
phenomenon.
\end{enumerate}

\section*{Acknowledgements}
\indent \indent I am very thankful to Timo Laitinen for his help on
the paper and Oskari Saloniemi for the editing of the anisotropy.
This work has been funded by grant from $Oskar$ $\ddot{O}flund$
$S\ddot{a}\ddot{a}ti\ddot{o}$.

\newpage

\textbf{Figure 1.}Types of the MESEP events according to the
intensity-time profile. From upper left to right, type 1 and 2, from
down left to right 3 and 4. Energy channels in MeV are listed on
each channel. The vertical lines are : dashed for associated solar
flares and arrow head for the associated CMEs. \vspace{10ex}

\textbf{Figure 2.} The event of 11.11.2000, Anisotropy and He/p
ration fit with a hidden CME during the time of the propagation of
particles acceleration. Energy channels in MeV are listed on each
channel. For explanation of the anisotropy diagrams, see text.
\vspace{10ex}

Figures are available on YSC home page
(http://ysc.kiev.ua/abs/proc14$\_$1.pdf).

\begin{table*}
\begin{center}
\section*{Appendix}
\caption{List of MESEP events during solar cycle 23}
\label{table1} \scriptsize{
\begin{tabular}{lccccccccccc}
    \hline \\
    Event & & & \multicolumn{3}{c}{ Associated eruptions } & &\multicolumn{2}{c}{ Type of event according to} \\
    \cline{1-2} \cline{4-6} \cline{8-9} \\
No & D.M.Y&&CME &Solar flare &&&energy level&Intensity-time profile \\
\hline\\

01&12-15.07.96&&4&2C&&&Mid&1\\
02&04-09.11.96&&2&--&&&Low&1\\
03&27-28.11.96&&2&--&&&Mid&2\\
04&28.11-03.12.96&&3&2C,1M&&&High&2\\
05&07-08.02.97&&3&--&&&Low&1\\
06&01-06.04.97&&3&1C&&&High&4\\
07&07-12.04.97&&2&1C&&&High&1\\
08&12-17.05.97&&2&--&&&High&1\\
09&21-24.05.97&&2&1M&&&High&1\\
10&20-22.09.97&&3&1C&&&High&1\\
11&24-29.09.97&&7&2M&&&High&1\\
12&12-14.10.97&&2&--&&&Mid&1\\
13&21-24.10.97&&5&1C&&&High&1\\
14&03-13.11.97&&8&2X,5M,1C&&&High&3\\
15&13-16.11.97&&5&1M&&&High&3\\
16&06-10.12.97&&4&--&&&High&1\\
17&03-04.01.98&&1&1M&&&low&1\\
18&25-31.01.98&&5&1C,1M&&&High&1\\
19&20-26.04.98&&6&1C,1M,1X&&&High&4\\
20&26.04-02.05.98&&5&3M,1X&&&High&2\\
21&02-06.05.98&&4&1C,2M,1X&&&High&4\\
22&06.05.98&&2&1M,1X&&&High&2\\
23&09-10.05.98&&3&3M&&&High&1\\
24&30-31.05.98&&3&1C&&&High&2\\
25&02-03.06.98&&7&1C&&&Mid&3\\
26&04-08.06.98&&6&--&&&High&1\\
27&08-14.06.98&&17&4C,2M&&&High&4\\
28&14-16.06.98&&3&1C&&&Mid&1\\
29&16-22.06.98&&13&1C,1M&&&High&1\\
30&22-24.06.98&&4&--&&&Mid&2\\
31&20-23.10.98&&2&1C&&&High&1\\
32&25-27.10.98&&7&--&&&Mid&4\\
33&05-13.11.98&&15&2C,9M&&&High&3\\
34&22-30.11.98&&9&4M,5X&&&High&3\\
35&09-13.12.98&&6&1C&&&Low&4\\
36&17-18.12.98&&5&1M&&&Mid&4\\
37&19-21.12.98&&2&1M&&&High&1\\
38&08-14.02.99&&8&3C,2M&&&High&1\\
39&25-26.02.99&&3&--&&&Mid&1\\
40&05-15.03.99&&21&1C,2M&&&Mid&1\\
41&25-27.03.99&&3&--&&&High&1\\
42&24-30.04.99&&13&1M&&&High&1\\
43&03-27.05.99&&25&2C,14M&&&High&3\\
44&27-30.05.99&&4&1C&&&High&4\\
45&01-03.06.99&&8&1C,1M&&&High&1\\
46&03-06.06.99&&5&1M6&&&High&4\\
47&11.06.99&&2&1C&&&High&1\\
48&24-27.06.99&&6&1M&&&Mid&2\\
49&27-28.06.99&&7&1C,1M&&&High&4\\
50&29.06-03.07.99&&9&11M&&&High&4\\
51&03-06.07.99&&6&1M&&&Mid&1\\
52&12-13.07.99&&3&--&&&Mid&1\\
53&16-18.07.99&&2&1C,2M&&&High&1\\
54&25-31.07.99&&14&1C,7M&&&High&1\\
55&31.07-07.08.99&&14&8M,1X&&&High&1\\
56&07-13.08.99&&8&4C,2M&&&Mid&1\\
57&17-21.08.99&&8&5C,6M&&&Mid&4\\
58&28-31.08.99&&1&1C,2M,1X&&&High&4\\
59&02-04.09.99&&3&2C&&&Mid&2\\
60&10-13.09.99&&9&--&&&Mid&1\\
61&13-16.09.99&&17&2C&&&High&1\\
62&19-20.09.99&&5&1C&&&Mid&1\\
63&21-22.09.99&&6&1C&&&Mid&1\\
64&13-14.10.99&&2&--&&&Mid&4\\
65&14-22.10.99&&13&1M,1X&&&High&1\\
66&22-24.10.99&&2&1C&&&Mid&1\\
67&06-14.11.99&&12&3C,12M&&&Low&1\\
68&15-23.11.99&&6&1C,14M&&&High&4\\
69&28.11-03.12.99&&1&4M&&&High&4\\
70&09-14.12.99&&2&1C&&&Mid&3\\
71&20-26.12.99&&8&3M&&&High&4\\
\hline\\
\end{tabular}}
\end{center}
\end{table*}

\begin{table*}
\begin{center}
\scriptsize{
\begin{tabular}{lccccccccccc}
    \hline \\
    Event & & & \multicolumn{3}{c}{ Associated eruptions } & &\multicolumn{2}{c}{ Type of event according to} \\
    \cline{1-2} \cline{4-6} \cline{8-9} \\
No & D.M.Y&&CME &Solar flare &&&energy level&Intensity-time profile \\
\hline\\

72&27-29.12.99&&7&3M&&&High&1\\
73&30.12.99-02.01.00&&8&--&&&Mid&4\\
74&09-11.01.00&&--&2M&&&High&2\\
75&18-23.01.00&&11&4M&&&High&2\\
76&27-31.01.00&&9&--&&&High&1\\
77&09-16.02.00&&21&3C,1M&&&High&1\\
78&17-23.02.00&&14&1C,9M&&&High&1\\
79&04-08.04.00&&13&4M&&&High&1\\
80&18-23.04.00&&11&4M&&&High&2\\
81&23-26.04.00&&12&--&&&High&4\\
82&27-30.04.00&&7&3C&&&High&3\\
83&04-11.05.00&&18&3M&&&High&2\\
84&11-14.05.00&&11&2M&&&High&2\\
85&14-26.05.00&&31&2C,12M&&&High&2\\
86&03-13.06.00&&12&--&&&High&4\\
87&13-16.06.00&&30&2C,11M,3X&&&High&4\\
88&16-21.06.00&&12&1C,1M,1X&&&High&3\\
89&21-25.06.00&&8&3M&&&High&1\\
90&25-28.06.00&&10&2M&&&High&1\\
91&28-30.06.00&&5&--&&&High&1\\
92&11-28.07.00&&36&42M,3X&&&High&2\\
93&28.07-01.08.00&&12&--&&&High&4\\
94&09-11.08.00&&4&3C&&&High&1\\
95&12-20.08.00&&24&1C,1M&&&High&2\\
96&04-07.09.00&&4&1C,1M&&&Mid&3\\
97&07-11.09.00&&9&1M&&&High&3\\
98&12-30.09.00&&33&3C,12M,1X&&&High&1\\
99&10-15.10.00&&6&2C,2M&&&High&4\\
100&16-24.10.00&&8&4C,2M&&&High&4\\
101&29.10-08.11.00&&19&3C,2M&&&High&2\\
102&08-24.11.00&&39&1C,11M&&&High&4\\
103&24.11-13.12.00&&31&4C,7M,5X&&&High&1\\
104&28.12.00-05.01.01&&10&2C,3M&&&High&1\\
105&05-09.01.01&&7&2C&&&High&1\\
106&21-28.01.01&&8&3M&&&High&1\\
107&28.01-06.02.01&&11&3C,2M&&&High&2\\
108&11-21.02.01&&11&1C&&&High&1\\
109&26.02-05.03.01&&14&2C&&&High&4\\
110&08-14.03.01&&20&2C,4M&&&High&3\\
111&25-29.03.01&&11&3C,11M&&&High&1\\
112&29.03-02.04.01&&15&14M,1X&&&High&1\\
113&02-09.04.01&&25&1C,11M,5X&&&High&3\\
114&09-15.04.01&&15&1C,5M,2X&&&High&3\\
115&15-25.04.01&&26&13M,1X&&&High&2\\
116&26.04-05.05.01&&11&2C,5M&&&High&2\\
117&07-20.05.01&&27&5C,5M&&&High&1\\
118&20-26.05.01&&12&3C,3M&&&High&4\\
119&26-30.05.01&&11&--&&&Mid&1\\
120&31.5-11.06.01&&27&4M&&&High&1\\
121&11-30.06.01&&43&5C,9M,1X&&&High&3\\
122&09-31.08.01&&50&3C,18M,1X&&&High&1\\
123&04-09.09.01&&6&1C,15M&&&Mid&4\\
124&12-14.09.01&&5&1C,2M&&&High&1\\
125&15-17.09.01&&4&7M&&&High&4\\
126&18-23.09.01&&7&5M&&&High&1\\
127&24-30.09.01&&17&1C,12M,1X&&&High&3\\
128&01-08.10.01&&17&5C,5M&&&High&3\\
129&09-18.10.01&&19&2C,2M&&&High&2\\
130&19-21.10.01&&5&8M,2X&&&High&4\\
131&22-31.10.01&&29&3C,16M,2X&&&High&1\\
132&02-04.11.01&&1&1M&&&Low&4\\
133&04-16.11.01&&14&2C,29M,1X&&&High&3\\
134&17-21.11.01&&4&1M&&&High&3\\
135&22-30.11.01&&12&1C,12M,1X&&&High&3\\
136&09-24.12.01&&23&2C,15M,2X&&&High&2\\
137&26.12.01-09.01.02&&23&3C,23M,1X&&&High&2\\
138&10-20.01.02&&13&3C,11M&&&High&3\\
139&27-31.01.02&&7&1C,1M&&&High&4\\
140&14-17.02.02&&3&--&&&Mid&2\\
141&19-28.02.02&&14&4C,12M&&&High&1\\
142&04-11.03.02&&12&2C,2M&&&Mid&3\\
143&11-22.03.02&&27&1C,9M&&&High&3\\
144&22-26.03.02&&7&2C,1M&&&High&3\\
145&30.03-02.04.02&&10&3M&&&Low&4\\
146&11-13.04.02&&7&2M&&&High&4\\
147&14-15.04.02&&6&4M&&&High&3\\
\hline\\
\end{tabular}}
\end{center}
\end{table*}
%
\begin{table*}
\begin{center}
\scriptsize{
\begin{tabular}{lccccccccccc}
    \hline \\
    Event & & & \multicolumn{3}{c}{ Associated eruptions } & &\multicolumn{2}{c}{ Type of event according to} \\
    \cline{1-2} \cline{4-6} \cline{8-9} \\
No & D.M.Y&&CME &Solar flare &&&energy level&Intensity-time profile \\
\hline\\

148&17-20.04.02&&7&1M&&&High&1\\
149&21.04-15.05.02&&50&1C,4M,1X&&&High&2\\
150&15-20.05.02&&27&2C,6M&&&Low&2\\
151&20-22.05.02&&3&1C,1M,1X&&&High&2\\
152&22-27.05.02&&15&2C,1M&&&High&4\\
153&27.05-06.06.02&&14&4C,7M&&&High&3\\
154&06-11.06.02&&7&2C&&&Mid&4\\
155&04-06.07.02&&6&2M&&&High&3\\
156&07-14.07.02&&15&1C,7M&&&High&1\\
157&16-31.07.02&&27&4C,16M,3X&&&High&2\\
158&03-08.08.02&&14&1C,3M,1X&&&High&3\\
159&14-16.08.02&&8&4M&&&High&1\\
160&16-21.08.02&&22&1C,16M,1X&&&High&3\\
161&22-24.08.02&&8&1C,7M&&&High&4\\
162&24-31.08.02&&17&11M,2X&&&High&1\\
163&05-16.09.02&&21&3C,6M&&&High&4\\
164&17-19.09.02&&5&1C&&&Mid&1\\
165&23.09.02&&2&--&&&Low&1\\
166&24.09.02&&4&--&&&High&1\\
167&27-30.09.02&&12&4M&&&High&1\\
168&05.10.02&&1&2M&&&Low&1\\
169&13-25.10.02&&32&2C,8M&&&High&1\\
170&28.10-06.11.02&&34&2C,4M,1X&&&High&4\\
171&09-17.11.02&&22&2C,8M&&&High&2\\
172&17-18.11.02&&2&2C,2M&&&Low&4\\
173&21-24.11.02&&7&--&&&Low&2\\
174&25.11-02.12.02&&24&3C&&&High&1\\
175&08-18.12.02&&14&6M&&&High&2\\
176&19-21.12.02&&6&1C,2M&&&High&4\\
177&22-25.12.02&&2&1M&&&High&2\\
178&30.12.02-03.01.03&&7&--&&&Low&1\\
179&20-23.01.03&&14&1C,2M&&&High&1\\
180&23-25.01.03&&4&3M&&&Low&1\\
181&27-30.01.03&&5&1C&&&High&1\\
182&01-03.02.03&&4&--&&&Low&1\\
183&12-13.02.03&&6&--&&&Low&1\\
184&20-23.02.03&&5&2C&&&Low&1\\
185&23-26.02.03&&4&--&&&Mid&4\\
186&13-17.03.03&&5&--&&&Low&1\\
187&17-18.03.03&&6&2M,1X&&&High&3\\
188&18-20.03.03&&10&5M,1X&&&High&2\\
189&24-30.03.03&&8&1C&&&Low&2\\
190&01-06.04.03&&15&1C,1M&&&Mid&4\\
191&07-12.04.03&&7&1C,1M&&&High&4\\
192&21-25.04.03&&7&4M&&&High&2\\
193&25.04-02.05.03&&20&6C,8M&&&High&4\\
194&06-09.05.03&&5&2C&&&Low&1\\
195&27.05-03.06.03&&20&3C,14M,3X&&&High&3\\
196&05-07.06.03&&11&1M&&&Mid&4\\
197&09-11.06.03&&4&11M,2X&&&Mid&2\\
198&11-13.06.03&&1&14M,1X&&&High&1\\
199&15-28.06.03&&14&5C,3M,1X&&&High&1\\
200&10-12.07.03&&4&2C,2M&&&High&1\\
201&17-21.07.03&&5&--&&&High&4\\
202&26.07-02.08.03&&6&1C,3M&&&Mid&1\\
203&03-07.08.03&&6&1M&&&Mid&2\\
204&08-10.08.03&&3&1C&&&Low&1\\
205&19-21.08.03&&4&2M&&&High&2\\
206&21-24.08.03&&4&1C&&&Low&4\\
207&17-23.09.03&&11&3C&&&Mid&1\\
208&04-05.10.03&&8&1M&&&High&2\\
209&20-26.10.03&&24&2C,2M,3&&&High&1\\
210&26-28.10.03&&6&6M,1X&&&High&1\\
211&28-29.10.03&&1&1X&&&High&3\\
212&29.10-02.11.03&&5&1C,9M,1X&&&High&1\\
213&02-04.11.03&&8&3M,3X&&&High&3\\
214&04-18.11.03&&28&1C,11M,1X&&&High&2\\
215&18-30.11.03&&21&2C,7M&&&High&2\\
216&02-11.12.03&&8&1C,4M&&&High&3\\
217&11-23.12.03&&12&--&&&Mid&2\\
218&26-27.12.03&&2&1M&&&Low&1\\
219&01-18.01.04&&30&1C,8M&&&High&1\\
220&18-28.01.04&&17&1C,4M&&&Mid&2\\
221&01-03.02.04&&4&--&&&Low&2\\
222&04-06.02.04&&9&--&&&High&4\\
223&26.02-03.03.04&&4&1M,1X&&&Mid&4\\
224&04-11.03.04&&8&1M&&&Low&2\\

\hline\\
\end{tabular}}
\end{center}
\end{table*}

\begin{table*}
\begin{center}
\scriptsize{
\begin{tabular}{lccccccccccc}
    \hline \\
    Event & & & \multicolumn{3}{c}{ Associated eruptions } & &\multicolumn{2}{c}{ Type of event according to} \\
    \cline{1-2} \cline{4-6} \cline{8-9} \\
No & D.M.Y&&CME &Solar flare &&&energy level&Intensity-time profile \\
\hline\\
225&17-19.03.04&&3&1C,2M&&&Low&2\\
226&21-23.03.04&&3&1C&&&Mid&2\\
227&08-17.04.04&&24&--&&&Low&2\\
228&20-25.05.04&&6&--&&&Low&4\\
229&25-28.05.04&&5&--&&&Low&4\\
230&02-04.06.04&&5&--&&&Mid&1\\
231&04-06.06.04&&3&--&&&High&4\\
232&06-17.07.04&&10&1C,9M,5X&&&High&1\\
233&22-25.07.04&&11&9M&&&High&1\\
234&25-29.07.04&&7&8M&&&High&2\\
235&29.7-06.08.04&&7&3C&&&High&3\\
236&13-17.08.04&&2&17M,1X&&&Mid&4\\
237&18-22.08.04&&7&1C,2M,1X&&&High&4\\
238&03-05.09.04&&7&--&&&High&4\\
239&09-19.09.04&&10&3C,3M&&&High&1\\
240&19-25.09.04&&5&1C,1M&&&High&4\\
241&30-31.10.04&&7&M,1X&&&High&2\\
242&01-03.11.04&&4&1C,5M&&&High&4\\
243&05-19.11.04&&24&8M,2X&&&High&1\\
244&03-12.12.04&&11&1C&&&High&1\\
245&15.01-15.02.05&&53&16M,5X&&&High&1\\
246&17-21.02.05&&7&1M&&&High&1\\
247&26.02-05.03.05&&6&--&&&Mid&2\\
248&06-10.03.05&&6&--&&&Mid&2\\
249&14-16.03.05&&2&1C&&&Mid&3\\
250&03-13.05.05&&19&1C,7M&&&High&1\\
251&13-20.05.05&&7&5M&&&High&3\\
252&26-29.05.05&&4&1C,1M&&&Mid&1\\
253&31.5-03.06.05&&3&3M&&&High&1\\
254&03-11.06.05&&6&--&&&High&4\\
255&16-24.06.05&&7&1C,1M&&&High&1\\
256&07-12.07.05&&17&3C,2M&&&High&3\\
257&13-22.07.05&&25&2C,9M,1X&&&High&2\\
258&24.7-07.08.05&&24&3C,7M,1X&&&High&4\\
259&22-28.08.05&&27&1C,4M&&&High&3\\
260&29-31.08.05&&8&1C&&&High&4\\
261&04-24.09.05&&27&4C,26M,10X&&&High&2\\
262&29.11-03.12.05&&12&5M&&&Mid&3\\
263&18-20.12.05&&3&--&&&Low&2\\
264&06-13.07.06&&12&1M&&&High&2\\
265&16-21.08.06&&6&2C&&&High&1\\
266&01-05.09.06&&7&1C&&&Mid&1\\
267&06-16.11.06&&13&1C&&&High&2\\
268&05-23.12.06&&19&4C,5M,4X&&&High&3\\
\hline\\
\end{tabular}}
\end{center}
\end{table*}


\begin{thebibliography}{27}
{\small
\bibitem{Al06}Al-Sawad A., Torsti J., Kocharov L., Huttunen-Hiekinmaa K.
JGR, V. 111, A10S90 (2006)
\bibitem{bom06}Bombardieri D.J., Duldig M.L., Michael K.J., Humble J.E. ApJ, V. 644, p. 565-574 (2006)
\bibitem{can86}Cane H.V., McGuire R.E., von Rosenvinge T.T. ApJ, V. 301,
p. 488-459 (1986)
\bibitem{can03}Cane H.V., von Rosenvinge T.T., Cohen C.M.S., Mewaldt
R.A. GRL, V. 30, 12, pp. SEP 5-1 (2003)
\bibitem{can02}Cane H.V., Erickson W.C., Prestage N.P. JGR, V. 107, A10, pp. 1-14 (2002)
\bibitem{cli96}Cliver E.W. in Ramaty R., Mandzhavidze N., X.-M. Hua (eds),
High Energy Solar Physics, AIP Conf. Proc 374, AIP press, Woodbury,
NY, p. 45 (1996)
\bibitem{cli04}Cliver E.W., Kahler S.W., Reames D.V. ApJ, V. 605, pp. 902-910 (2004)
\bibitem{jia06a}Jian L., Russell C.T., Luhmann J.G., Skoug R. M. Solar
Physics, V. 239, 1-2, pp. 337-392 (2006)
\bibitem{jia06b}Jian L., Russell C.T., Luhmann J.G., Skoug R.M. Solar
Physics, V. 239, 1-2, pp. 393-436 (2006)
\bibitem{kah77}Kahler S.W. ApJ, V. 214, pp. 891-897 (1977)
\bibitem{kah94}Kahler S. ApJ, V. 428, pp. 837-842 (1994)
\bibitem{kah01}Kahler S. JGR, V. 106, A10, pp. 20947-20955 (2001)
\bibitem{koc02}Kocharov L., Torsti J. Solar Physics, V. 207, pp. 149-157 (2002)
\bibitem{rea99}Reames D.V. Space Sci. Rev., V. 90(3/4), pp. 413-491 (1999)
\bibitem{she75}Sheeley N.R. Jr., Bohlin J.D., Brueckner G.E., et al. Solar Physics, V. 45, pp. 377-392 (1975)
\bibitem{she83a}Sheeley N.R. Jr., Howard R.A., Koomen M.J., Michels D.J. ApJ, V. 272, pp. 349-354 (1983a)
\bibitem{she83b}Sheeley N.R. Jr., Howard R.A., Koomen M.J., et al. JPL Solar Wind Five pp. 693-702 (1983b)
\bibitem{she85}Sheeley N.R. Jr., Howard R.A., Michels D.J., et al. JGR (ISSN 0148-0227), V. 90, pp. 163-175 (1985)
\bibitem{tor97}Torsti J., Laitinen T., Vainio R., et al.  Solar Physics, V. 175, pp. 771-784 (1997)
\bibitem{tor99}Torsti J., Kocharov L., Teittinen M., et al. JGR, V. 104, pp. 9903-9910 (1999)
\bibitem{tor04}Torsti J., Riihonen E., Kocharov L. ApJ, V. 600, pp. L83-L86 (2004)
\bibitem{tor06}Torsti J., M$\ddot{a}$kel$\ddot{a}$ P., Riihonen E., Saloniemi O. ApJ, V. 638, pp. 530-518 (2006)
}
\end{thebibliography}
\end{document}